\documentclass[10pt]{article}
\usepackage[pctex32]{graphics}
\usepackage[dvips]{graphicx}

\usepackage{epsfig}
\usepackage{graphicx}
\usepackage{indentfirst}
\usepackage{amsmath}
\usepackage{amsfonts}
\usepackage{amssymb}
\usepackage{hyperref}
\usepackage{latexsym}
\usepackage{color}

\begin{document}
\begin{center}
{\Large\bf Stability and Space Phase Analysis in $f(R)$ theory with Generalized Exponential model}\\

\medskip

R. D. Boko$^{(a)}$\footnote{e-mail:docenzi@yahoo.fr}, M. J. S. Houndjo$^{(a,b)}$\footnote{e-mail:
sthoundjo@yahoo.fr} 
 and J. Tossa$^{(a)}$\footnote{e-mail: joel.tossa@imsp-uac.org}

$^a$ \,{\it Institut de Math\'{e}matiques et de Sciences Physiques (IMSP)}\\
 {\it 01 BP 613,  Porto-Novo, B\'{e}nin}\\

$^{b}$\,{\it Facult\'e des Sciences et Techniques de Natitingou - Universit\'e de Natitingou - B\'enin} \\

\date{}

\end{center}
\begin{abstract}
We have studied in this paper, the stability of dynamical system in $f(R)$ gravity. We have considered the $f(R)$ $\gamma$-gravity
 and explored its dynamical analysis. We found six critical points among which only one 
describes an universe fulled of both matter and dominated dark energy. It's  shown that these critical points presents specific phase 
spaces described by the corresponding fluids. Furthermore, we've investigated the stability conditions of these critical points and find 
that theses conditions are dependent of the model parameters. We also study the stability of a new power-law $f_\ast(R)$ model with de Sitter and power law solutions.
\section{Introduction}
The cause of the recent acceleration of universe has been discovered through the observations data of the supernovae-type Ia (SNe Ia),
the large-scale structure (LSS) and the cosmic microwave background radiation (CMBR): it's an exotic fluid with negative pressure named
dark energy. Since $\Lambda$CDM model met some limits to explain correctly the cosmic acceleration, many searchers have tried to explain
this acceleration by introducing different alternatives theories in the Einstein-Hilbert action. Among these theories,
we have $f(R)$, $f(G)$, $f(R,G)$, $f(R,\mathcal{T})$, $f(T,\mathcal{T})$ modified gravity where $R$ is the Ricci scalar,
$G$ the gravity constant and $\mathcal{T}$  is the trace of the energy-momentum tensor. Many authors have explored $f(R)$ theory 
in different ways in order to show the properties of such a fluid. V. Linder et al. has investigated a exponential $f(R)$ gravity 
to explain cosmic acceleration \cite{Rach13}-\cite{Rach16}. In this work, they calculate the dynamics and examine 
the power spectrum effect. Exponential $f(R)$ model has been investigated by some authors in different ways \cite{Rach9}-\cite{Rach12} 
.Bamba et al. have explored the cosmological evolution within $f(R) = R + c_1(1-e^{-c_2 R})$ $(c_{1,2}$ = constant)  exponential gravity. 
In that paper, they have shown that the exponential gravity can be considered as a viable gravitational modified theory through
its viability conditions. It's also shown that the late-time cosmic acceleration can be realized after the matter-dominated stage.\\ 
Recently, it has been presented a generalization of exponential $f(R)$ gravity theory named $\gamma$ gravity
\Big[$f(R)=-\frac{\alpha R_\ast}{n}\gamma\left(\frac{1}{n},(\frac{R}{R_\ast})^n\right)$\Big] where $\alpha$, $R_\ast$ and $n$ are 
the positive parameters \cite{Rach18}: the effective equation of state parameter behavior has been examined as a redshifts function
for different values of the parameters $\alpha$ and $n$, the dynamical analysis of $f(R)$ and $f(T)$ models has been 
explored and these results gave one parameter more than the one of General Relativity. This parameter depends on the dark energy model,
it gives the stability conditions for the critical points \cite{Rach17},\cite{Rach35}.
Many authors has explored the phase space and study the stability of critical points with some $f(R)$-Lagrangian model and presented the important
results \cite{Rach41}-\cite{Rach47}. \\ In this paper,following the method of \cite{Rach41}, we explore the $R+f(R)$ gravity 
theory in the same way and obtain two critical points. These points live on the spaces phases defined by the dynamics fluid of the universe. 
The discussion around the critical points stability yields some conditions that must be fulfilled by the parameters of the generalized exponential
$f(R)$ model.In the following we defined a power-law model in $f(R)$ gravity taking the form: $(a(b+c(d+R)^p)$, for $p=-1$. We shown that 
this function have the same behavior like the one of $ f(R)\gamma$-gravity model .After, we proceed to the stability analysis of that model which gives important
results with de Sitter and power law solutions.\\ The paper is organized as follow: After establishing both Friedmann equations
with $R+f(R)$ theory in the second section, we established in the third section, the dynamical system using in the action the form $ R+f(R)$. 
We then study the stability of dynamical system with $f(R)$ $\gamma$-gravity in the fourth section. The five section is devoted to the stability
analyze of the generalized power-law $f(R)$ model and the paper is ended by the conclusion in section five.

\section{Motion Equations within $R+f(R)$ modified gravity theory}
\end{abstract}
Here, we consider the Einstein-Hilbert action in which the Ricci scalar $R$ is added to a $f(R)$ function and the action is writing as:
\begin{eqnarray}
S=\int dx^4\sqrt{-g}\left\{\frac{1}{2k^2}\Big[R+f(R)\Big]+L_m\right\},
\end{eqnarray}
where $g$ is the determinant of the metric $g_{\mu\nu}$, $\kappa^2\equiv8\pi G$ with $G$ the gravitational constant and  $L_m$ the mater Lagrangian. The general equation of motion in $R+f(R)$ modified gravity is given by:
\begin{eqnarray}
(1+f_R)R_{\mu\nu}-\frac{1}{2}g_{\mu\nu}(R+f(R))+(g_{\mu\nu}\square-\nabla_\mu\nabla_\nu)f_R=k^2T_{\mu\nu}      \label{R0}
\end{eqnarray}
Considering temporal components, the equation ($\ref{R0}$) give the first and second equation of Friedmann:
\begin{eqnarray}
H^2+\frac{f}{6}-f_R(H^2+\dot{H})+H\dot{R}f_{RR}=\frac{k^2}{3}\rho          \label{R1}
\end{eqnarray}
\begin{eqnarray}
(\ddot{R}-H\dot{R})f_{RR}+2\dot{H}(1+f_R)+\dot{R}^2f_{RRR}=-k^2(\rho+p)     \label{R2}
\end{eqnarray}
where $\rho$ and $p$ are respectively the energy density and the pressure of the total fluid of universe and are given by: 
$\rho=\rho_m+\rho_r$ and $p=p_m+p_r$. The indication $m$ means ordinary mater and $r$ the radiation.\\
One define: $f_R\equiv\frac{df(R)}{dR}$ , $f_{RR}\equiv\frac{d^2f(R)}{dR^2}$ and $f_{RRR}\equiv\frac{d^3f(R)}{dR^3}$.
In the following, we will adopt the writing $F$ instead of $f_R$.\\
From equations (\ref{R1}) , (\ref{R2}) we can obtain the pressure and density energy :
\begin{eqnarray}
 p_d=\frac{1}{2\kappa^2}[f-2F(\dot{H}+3H^2)+2(2H\dot{R}+\ddot{R})f_{RR}+2\dot{R}^2f_{RRR}                       \label{Rp}
\end{eqnarray}
\begin{eqnarray}
 \rho_d=\frac{1}{2\kappa^2}\Big[-f+F(R-6H^2)-6H\dot{R}f_{RR}\Big]                                                     \label{Rd}
\end{eqnarray}
Then, using the relations (\ref{Rp}) and (\ref{Rd}) we can write the dark energy equation of state as follow:
\begin{eqnarray}
 w_d=\frac{f-2F(\frac{R}{6}-H^2)+2(2H\dot{R}+\ddot{R})f_{RR}+2\dot{R}^2f_{RRR}}{-f+F(R-6H^2)-6H\dot{R}f_{RR}}              \label{Re}
\end{eqnarray}

\section{Dynamical system in general f(R) gravity }
The continuity equation for the total fluid of the universe is:
\begin{eqnarray}
\dot{\rho}+3H(\rho+p)=0                                    \label{Rcont}
\end{eqnarray}

It's discussed in \cite{Rach17} that the dynamical system within $f(R)$ gravity theory presents four equations with four variables. Carloni 
et. al have explored a autonomous dynamical system   some of the critical points found depends on a parameter which depends on the $f(R)$ model. \\
The equation (\ref{R1}) can be writing as:
\begin{eqnarray}
1=\frac{-f}{6H^2}+\frac{FR}{6H^2}-F-\frac{\dot{F}}{H}+\frac{\kappa^2\rho}{3H^2}            \label{R5}
\end{eqnarray}
Let's introduce the dimensionless parameters $u_1$, \;$u_2$, \;$u_3$, \;$u_4$\; and \;$u_5$ defined by:
\begin{eqnarray}
u_1\equiv\frac{-f}{6H^2},\;\;\; u_2\equiv\frac{FR}{6H^2},\;\;\; u_3\equiv-F,\;\;\; u_4\equiv-\frac{\dot{F}}{H}\;\;et\;\; u_5\equiv\frac{\kappa^2\rho}{3H^2}
\end{eqnarray}

By using conservation equation (\ref{Rcont}) and the Friedmann equation (\ref{R5}), we obtain the dynamical system defined as follow:
\begin{eqnarray} \label{R6} 
\frac{du_1}{dN}&=&-\frac{u_2u_4}{u_3m}+2\frac{u_1u_2}{u_3}+4u_1 \\                            
\frac{du_2}{dN}&=&\frac{u_2u_4}{u_3}+\frac{u_2u_4}{u_3m}+2\frac{u_{2}^2}{u_3}+4u_2  \\    \label{R7}
\frac{du_3}{dN}&=&u_4     \\                                                              \label{R8}
\frac{du_4}{dN}&=&-2\frac{u_1u_2}{u_3}-\frac{u_2u_4}{u_3}-2\frac{u_2u_5}{u_3}-2\frac{u_2^2}{u_3}-4u_1-4u_2-u_4+u_5(3w-1) \\               \label{R9}
\frac{du_5}{dN}&=&2\frac{u_2u_5}{u_3}+u_5(1-3w)                                                   \label{R10}
\end{eqnarray}
where $N$ is the e-folding parameter $N\equiv lna$ and\\
\begin{eqnarray}
m\equiv\frac{R}{F}f_{RR}.
\end{eqnarray}
The $m$ parameter is defined as its inversion $q$ in \cite{Rach41}.
Then, following \cite{Rach41}, one can write $m$ in terms of the dynamical variables to perform the phase space analysis. So, we define an 
additional parameter $r$ in terms of the dynamical variables as:
\begin{eqnarray}
 r\equiv-\frac{RF}{f}=\frac{u_2}{u_1}
\end{eqnarray}
One can see that $m$ parameter could be in terms of $r$ and consequently in terms of dynamical variables.\\
This system is divergent for $u_3=0$ or $m=0$. This means that, finding attractors points will need some conditions.\\
Using $u_1$, \;$u_2$, \;$u_3$, \;$u_4$\; and \;$u_5$, the equation of state $w_d$ (\ref{Re}) take the form:
\begin{eqnarray}
 w_d=\frac{u_3[u_1+3u_2+u_4-u_3+u_5(1-3w)]+2u_2(1-u_3)}{3u_3(u_1+u_2+u_3+u_4)}
\end{eqnarray}
\section{Stability analysis of dynamical system with Gamma gravity f(R) model}
we are interested on $f(R)$ Gamma model which is a generalization of exponential model $f(R)$. It's given by :
\begin{eqnarray}
 f(R)=-\frac{\alpha R_\ast}{n}\gamma\left(\frac{1}{n},(\frac{R}{R_\ast})^n\right)                    \label{Rgam}
\end{eqnarray}
where $\gamma$ $(\alpha;z):=\int_0^ze^{-t}t^{\alpha-1}dt$ is the $\gamma$-function, with $\alpha$, $n$ and $R_\ast$ the positives parameters.\\
Many studies have been done with this model \cite{Rach18}, among some of these studies, one has the universe expansion history. The behavior of parameter of state is been examined as a redshift function for different values of parameters $\alpha$ and $n$ .
Also the constraints from local gravity tests and linear growth of structure was been discussed. In this paper, we are going to analyze the 
phase space of the dynamical system with the Lagrangian  gamma $f(R)$. Then following \cite{Rach41}, one express $m$ in terms of dynamical 
variables defined in above. With the relation (\ref{R5}) the function $m= m(u_1,u_2,u_4,u_5)$ is reduced to the $m=m(u_3)$ one, and is giving by:
\begin{eqnarray}
 m(u_3)=n(\ln u_3-\ln\alpha)                                                                        \label{Rgamu3}
\end{eqnarray}
Now, substituting (\ref{Rgamu3}) into the system (\ref{R6})-(\ref{R10}), one get
\begin{eqnarray} \label{R6syst2} 
\frac{du_1}{dN}&=&-\frac{u_2u_4}{u_3n(\ln u_3-\ln\alpha)}+2\frac{u_1u_2}{u_3}+4u_1 \\                            
\frac{du_2}{dN}&=&\frac{u_2u_4}{u_3}+\frac{u_2u_4}{u_3n(\ln u_3-\ln\alpha)}+2\frac{u_{2}^2}{u_3}+4u_2  \\    \label{R7syst2}
\frac{du_3}{dN}&=&u_4     \\                                                              \label{R8syst2}
\frac{du_4}{dN}&=&-2\frac{u_1u_2}{u_3}-\frac{u_2u_4}{u_3}-2\frac{u_2u_5}{u_3}-2\frac{u_2^2}{u_3}-4u_1-4u_2-u_4+u_5(3w-1) \\  \label{R9syst2}
\frac{du_5}{dN}&=&2\frac{u_2u_5}{u_3}+u_5(1-3w)                                                   \label{R10syst2}
\end{eqnarray}
As showing previously, the first and second equation of this system present a singularity on the point $u_3=\alpha$. With the gamma f(R)
function, this singularity is happened at the beginning $(R\rightarrow 0)$.
Now, let's search the critical points of the dynamical system (\ref{R6syst2})-(\ref{R10syst2}) by solving the equations: 
$\frac{du_1}{dN}=0$, \;$\frac{du_2}{dN}=0$, \; $\frac{du_3}{dN}=0$, \; $\frac{du_4}{dN}=0$, \;
$\frac{du_5}{dN}=0$. So doing, one obtains the critical points:
\begin{eqnarray}
&A_1&(u_1=u_1, \;u_2=-2u_3, \;u_3=u_3, \;u_4=0, \;u_5=0) , \\ 
&A_2&(u_1=0, \;u_2=0, \;u_3=u_3, \;u_4=0, \;u_5=0) , \\    
&A_3&(u_1=-u_2, \;u_2=u_2, \;u_3=\alpha,  \;u_4=0, \;u_5=0),  \\      
& A_4&(u_1=\frac{1}{2}(1-3w)\alpha, \;u_2=\frac{1}{2}(-1+3w)\alpha, \;u_3=\alpha, \;u_4=0, \;u_5=u_5),  \\  
 &A_5&(u_1=u_1, \;u_2=-2\alpha, \;u_3=\alpha, \;u_4=0, \;u_5=0),   \\        
& A_6&(u_1=0, \;u_2=0, \;u_3=\alpha, \;u_4=0, \;u_5=0) 
\end{eqnarray}
With the viable form $f(T)=\beta\sqrt{T}$, assuming that dark energy is a perfect fluid, it's shown that there is one attractor solution to the 
dynamical equation of $f(T)$ Friedmann equations. Also, studying the local stability near the critical points, it's shown that the critical points 
lie on the sheet $u^\ast=(c-1)v^\ast$ in the phase space, spanned by four coordinates \cite{Rach19}.\\
 Besides, $A_4$ fixed point, all the others fixed points correspond to an universe with no matter and only filled with the vacuum energy.\\
$ \bullet$ Point $A_1$: Dark energy dominated point\\
One can see that, this point lie on a two dimensional surface $u_2=-2u_3$, with a continuous presence of the dark energy. It present also a dark
energy line point on the axis $u_1$.\\
$\bullet$ Point $A_2$: Corresponding to a dark energy dominated universe\\
As previously with the above $A_1$ fixed point, the $A_2$ fixed point present a line $u_3$ in the phase space.\\
$\bullet$ Point $A_3$: Dark energy dominated point.\\
Here, the critical point $A_3$ lie on the surface $u_1=-u_2$ in the phase space. If $u_2=0$, then $u_1=0$, but $u_3=\alpha\neq 0$. Thus, the
fixed point $A_3$ corresponds to a full presence of the vacuum energy.\\
$\bullet$ Point $A_4$: Co-dominance of matter and dark energy.\\
This critical point correspond to an universe filled with matter and dark energy. For $w=\frac{1}{3}$(radiation fluid), the containing of the 
universe is both radiation and dark energy, but with $w=0$, $A_4$ shows an universe filled with dust and dark energy.\\
$\bullet$ Point $A_5$: Like $A_1$, $A_2$, $A_3$, the fixed point $A_4$ is the fourth dark energy dominated point of the dynamical system in
above. It describes an axis $u_1$ in the phase space.\\ 
$\bullet$ Point $A_2$: Dark energy dominated point.\\
We note that, this point doesn't present an space phase.

In order to study the stability of the critical points , we will consider the small perturbations around critical points
 and write each point of the system in the form $u_c+\delta u$. Thus, the previous dynamical system become a set of the 
following equations:
\begin{eqnarray}
\frac{d\delta u_1}{dN}&=&2\left(2+\frac{u_2}{u_3}\right)\delta u_1+\left(2\frac{u_1}{u_3}-\frac{u_4}{u_3[n(\ln u_3-\ln \alpha)}\right)\delta u_2-
\left(\frac{u_2u_4}{u_{3}^2[n(\ln u_3-\ln \alpha)]^2}+2\frac{u_1u_2}{u_3^2}\right)\delta u_3-\\&-&\frac{u_2}{u_3[n(\ln u_3-\ln \alpha)]}\delta u_4                                  \\
\frac{d\delta u_2}{dN}&=&\left(\frac{u_4}{u_3}+\frac{u_4}{u_3[n(\ln u_3-\ln \alpha)]}+4\frac{u_2}{u_3}+4\right)\delta u_2+\left(-\frac{u_2u_4}{u_3^2}+
\frac{u_2u_4}{u_3^2[n(\ln u_3-\ln \alpha)]^2}-2\frac{u_2^2}{u_3^2}\right)\delta u_3+\\&+&\left(\frac{u_2}{u_{3}}+\frac{u_2}{u_3[n(\ln u_3-\ln \alpha)]}\right)\delta u_4             \\
\frac{d\delta u_3}{dN}&=&\delta u_4                                                                                                 \\
\frac{d\delta u_4}{dN}&=& -2\left(\frac{u_2}{u_3}+2\right)\delta u_1-2\left(\frac{u_1}{u_3}+\frac{u_4}{2u_3}+\frac{u_5}{u_3}+2\frac{u_2}{u_3}+2\right)\delta u_2+
2\frac{u_2}{u_3}\left(\frac{u_1}{u_3}+\frac{u_4}{2u_3}+\frac{u_5}{u_3}+\frac{u_2}{u_3}\right)\delta u_3\\&-&
\left(\frac{u_2}{u_3}+1\right)\delta u_4+\left(3w-2\frac{u_2}{u_3}-1\right)\delta u_5                                              \\
 \frac{d\delta u_5}{dN}&=&2\frac{u_5}{u_3}\delta u_2-2\frac{u_2u_5}{u_3^2}\delta u_3+\left(2\frac{u_2}{u_3}+1-3w\right)\delta u_5
\end{eqnarray}
After some calculus, the eigenvalues of the critical points are respectively:\\
$\bullet$ Point $A_1$
\begin{eqnarray}
 \lambda_1&=&-2,\;\;\;\;\lambda_2=-3(1+w),\;\;\;\;\lambda_3=-\frac{1}{3}-[2^\frac{1}{3}(-(-nu_3\ln(\frac{u_3}{\alpha})^2+...+8nu_3\ln\alpha)]^\frac{1}{3},\;\;\;\;\\
 \lambda_4&=&-\frac{1}{3}+((1+i\sqrt{3}))[-(nu_3\ln(\frac{u_3}{\alpha})^2+....+8nu_3\ln\alpha)]^\frac{1}{3},\;\;\;\;\;\\
 \lambda_5&=&-\frac{1}{3}+((1-i\sqrt{3}))[-(nu_3\ln(\frac{u_3}{\alpha})^2+....+8nu_3\ln\alpha)]^\frac{1}{3}
\end{eqnarray}
$\bullet$ Point $A_2$
\begin{eqnarray}
 \lambda_6=-1,\;\;\;\;\lambda_7=0,\;\;\;\;\lambda_8=2,\;\;\;\;\lambda_9=4, \;\;\;\;\;\;\lambda_{10}=1-3w,
\end{eqnarray}
$\bullet$ Point $A_3$
\begin{eqnarray}
 \lambda_{11}=0,\;\;\;\;\lambda_{12}=2\frac{u_2}{\alpha}+3-u_2,\;\;\;\;\lambda_{13}=2\frac{u_2}{\alpha}+1-3w,
\end{eqnarray}
$\bullet$ Point $A_4$
\begin{eqnarray}
 \lambda_{14}=0,\;\;\;\;\lambda_{14}=0,\;\;\;\;\lambda_{15}=\frac{6u_5(1+w)}{2u_5+\alpha(1-3w)},
\end{eqnarray}
$\bullet$ Point $A_5$
\begin{eqnarray}
 \lambda_{16}=-2,\;\;\;\;\lambda_{17}=0,\;\;\;\;\lambda_{18}=-3(1+w),
\end{eqnarray}
$\bullet$ Point $A_6$
\begin{eqnarray}
 \lambda_{16}=-1,\;\;\;\;\lambda_{17}=0,\;\;\;\;\lambda_{18}=2,\;\;\;\;\lambda_{19}=4,\;\;\;\;\lambda_{20}=1-3w,
\end{eqnarray}
Firstly, one see that there is no global stable point, but some of them could be stable with some conditions on the parameters $w$, $\alpha$,
$u_2$, $u_3$ and $u_5$. One note that there are doubles fixed points: $A_2$ and $ A_6$.\\
The fixed point $A_1$ present two imaginary eigenvalues depending of the parameters $u_3$ and $\alpha$. It is an conditionally stable 
critical point. The fixed point $A_2$ is a saddle point for any value of $w$.\\ The fixed point $A_3$ is an attractor if 
$\frac{u_2}{\alpha}<\frac{3w-1}{2}$ and also $\frac{u_2}{\alpha}<\frac{\alpha}{\alpha-2}$. It will be considered as a saddle point if only 
one of the last condition is satisfied. $A_3$ should be a repellor if $\frac{u_2}{\alpha}>\frac{3w-1}{2}$ and $\frac{u_2}{\alpha}>\frac{\alpha}{\alpha-2}$.\\ 
Considering the fixed point $A_4$, its attractor point condition is $\frac{(1+w)}{2u_5+\alpha(1-3w)}<0$. Otherwise this condition, the 
fixed point  $A_4$ would be a reppellor.\\ The critical point $A_5$ is an attractor if $w>-1$, but a saddle otherwise. One can see that 
the fixed point $A_6$ present the same condition of stability with the fixed point $A_2$.\\
Note that, $A_4$ is an important critical point because it shown that with its stability condition, the universe could evolve from matter 
dominated phase to dark energy one.

\section{Stability Analysis of the $f_\ast(R)$ model}
In this section, we consider a type of power law f(R) having the same behavior like the exponential one. We focus our attention to its stability with de Sitter solution and the power-law solution.
\subsection{the $f_\ast(R)$ model behavior}
We consider the following ansatz,
\begin{eqnarray}
f_\ast(R)=-\alpha R_{\ast}\left[1-\frac{R_{\ast}}{\alpha(R_{\ast}+\beta R)}\right]                          \label{R16}   
\end{eqnarray}
where  $R_{\ast}$, $\alpha $ and $\beta$ are free positive parameters.  
 The dark energy f(R) model (\ref{R16}) is a generalization of the power-law f(R)($R+aR^p $) model for $p=-1$. Note that since $(\frac{R}{R_\ast}\gg 1)$, 
 General Relativity with cosmological constant is recovered in $R+f(R)$ gravity. Note that this function is defined, because $R\neq-\frac{R_\ast}{\beta}. $ 
 In the fig $1$ we plot the behavior of the $ f_\ast(R)$ models for different value of $\alpha$ and $\beta$.
  \newpage
\begin{figure}[h]
  \centering
  \begin{tabular}{rl}
  \includegraphics[height=5.5cm,width=8cm]{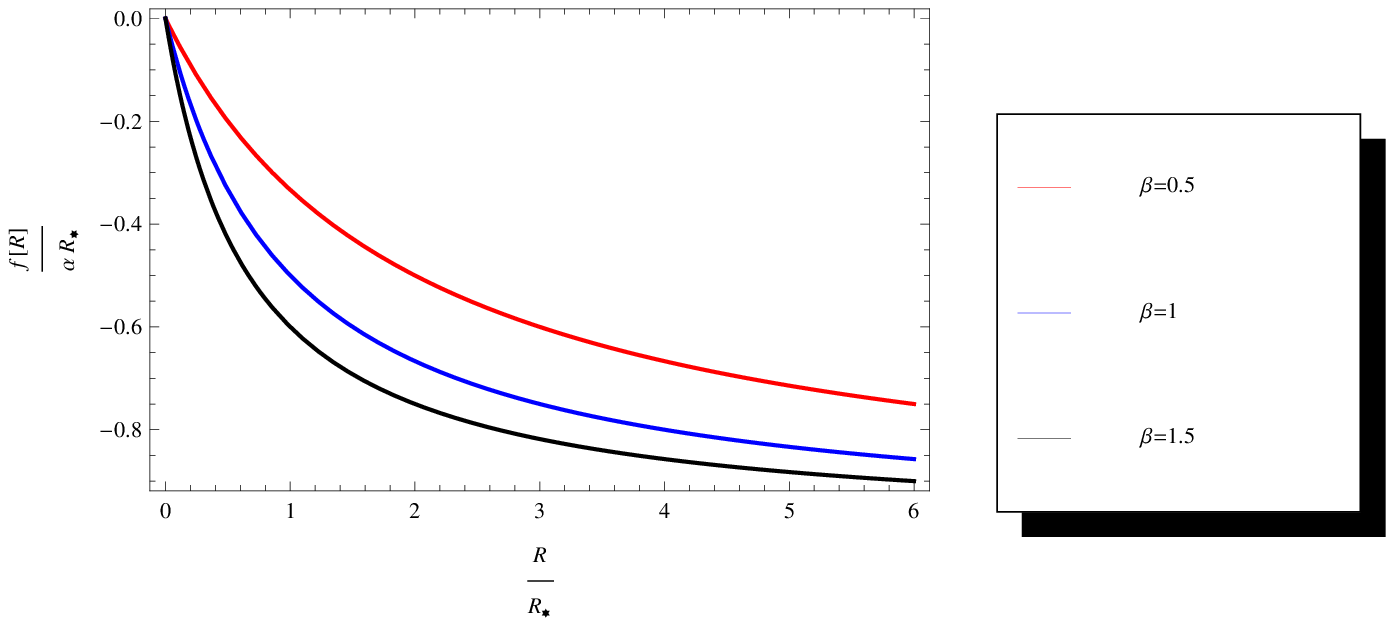}&
  \includegraphics[height=5.5cm,width=8cm]{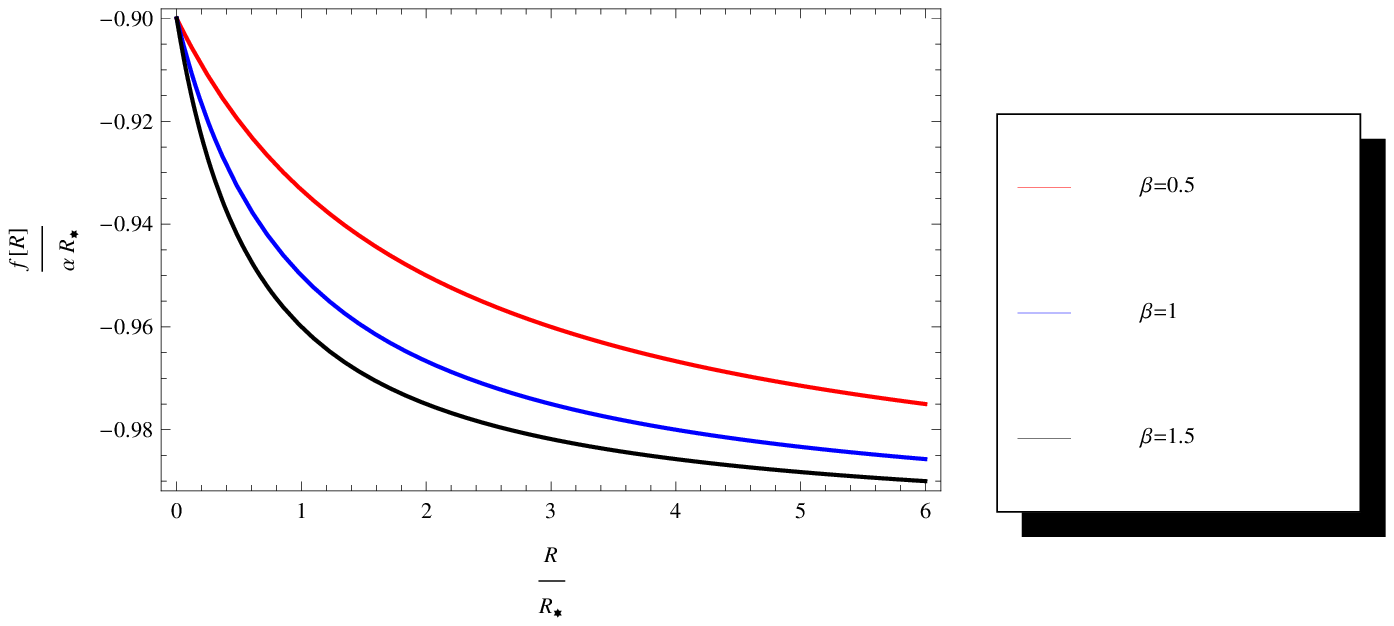} 
 \end{tabular}
 \caption{ The graph in the left-hand side shows the behavior of the cosmological model in terms of $\frac{R}{R_\ast}$, for 
 $\alpha=1$. The one in the right-hand side shows its evolution for $\alpha=10$.}
 \label{fig1}
 \end{figure}

Considering baryonic matter as universe fluid the solution of the equation (\ref{Rcont}) gives the matter density in the background:
\begin{eqnarray}
 \rho_{mb}(t)=\rho_{mo} e^{-3(1+w_m)\int H_b(t)dt}                                                            \label{R17}  
\end{eqnarray}
Now taking in account the Hubble parameter and the energy density of the matter in terms of perturbations:
\begin{eqnarray}
 H(t)=H_b(t)\Big[1+\delta(t)\Big]\;\;\;\;,\;\;\;\;\rho_m(t)=\rho_{mb}(t)\Big[1+\delta_m(t)\Big]                              \label{R18}  
\end{eqnarray}
where $H_b(t)$ and  $\rho_{mb}(t)$ are respectively the Hubble parameter and energy density of the baryonic matter at the background.\\
Let's develop the function $f(R)$  in a series of $R_b=12H_{b}^2(t)+6\dot{H}_b(t)$ as:
\begin{eqnarray}
f(R)=f^b(R_b)+f^{b}_R(R_b)(R-R_b)+ \frac{1}{2}f_{RR}^b(R_b)(R-R_b)^2+O^3.                             \label{R19}         \end{eqnarray}
Introduicing (\ref{R18}) in (\ref{R1}) and taking in account of (\ref{R19}) and of the expanding of its first and second derivatives, one obtain the following equation:
\begin{eqnarray} \label{R20}
6H_b^2(t)f_{RR}^b(R_b)\ddot{\delta}(t)+6H_b(t)f_{RR}^b(R_b)\Big[3H_b^2(t)+\dot{H}_b(t)\Big]\dot{\delta}(t)+               
2\Big[H_b^2(t)+H_b^2(t)f_R^b(R_b)
-12H_b^4(t)f_{RR}^b(R_b)+\nonumber\\+21H_b^2(t)\dot{H}_b(t)f_{RR}^b(R_b)-3\dot{H}_b^2(t)f_{RR}^b(R_b)+6H_b(t)\ddot{H}_b(t)f_{RR}^b(R_b)\Big]\delta(t)=\frac{\kappa^2}{3}\rho_{mb}(t)\delta_m(t)    
\end{eqnarray}
The relation in above shows the matter and geometry perturbation density. The resolution of this equation
need a second equation. It's then useful to consider the continuity equation (\ref{R3}) and making use of  (\ref{R18}), one obtain:
\begin{eqnarray}
\dot{\delta}_m(t)+3H_b(t)(1+\omega_m)\delta(t)=0                           \label{R21}     
\end{eqnarray}
We will proceed to the analysis of the perturbations functions $\delta_m(t)$ and $\delta(t)$ in some cases . Thus, with $\kappa^2=1$, one have:\\
$\bullet$ for large $R$ , this means that the high derivatives functions are negligible. Then, the equation (\ref{R20}) become:
\begin{eqnarray}
2H_b^2(t)\Big[1+f_R^b(R_b)\Big]\delta(t)=\frac{1}{3}\rho_{mb}(t)\delta_m(t)                               \label{R22}
\end{eqnarray}
$\bullet$ for small $R$ , and setting $\alpha=1$, the differential equation (\ref{R20}) is reduced to the General Relativity ones:
\begin{eqnarray}
2H_b^2(t)\delta(t)=\frac{1}{3}\rho_{mb}(t)\delta_m(t)                                                            \label{R23}
\end{eqnarray}
\subsection{Stability of de Sitter Solutions} 
The Hubble parameter in the background for the de Sitter solutions is a constant and is writing by:
\begin{eqnarray}
 H_b(t)=H_0\Longrightarrow a(t)=a_oe^{H_ot}                                                                     \label{R24}  
\end{eqnarray}
where $H_o$ is a constant, therefore (\ref{R17}) become:
\begin{eqnarray}
\rho_{mb}(t)=\rho_{mo} e^{-3(1+w_m) H_ot}                                                                        \label{R25}  
\end{eqnarray}
By introducing the de Sitter Hubble parameter expression (\ref{R24}) in the differential equations (\ref{R22}), (\ref{R23}) and taking care 
of (\ref{R21}) and (\ref{R25}), one find respectively:
\begin{eqnarray}
2H_0^2\Big[1+f_R^b(R_0)\Big]\dot{\delta}_m(t)=-H_0(1+w_m)\rho_{mo} e^{-3(1+w_m) H_ot}\delta_m(t)                \label{R26}
\end{eqnarray}
and
\begin{eqnarray}
2H_0^2\dot{\delta}_m(t)=-H_0(1+w_m)\rho_{mo} e^{-3(1+w_m) H_ot}\delta_m(t)                                     \label{R27}  
\end{eqnarray}
where $R_o=12H_o^2$ is the Ricci scalar of de Sitter on the background.\\
From the equations (\ref{R26}) and (\ref{R21}) one obtain, the perturbation density of the matter and of the  geometry:
\begin{eqnarray}
\delta_m(t)&=&k_1\exp\Big[\frac{\rho_{mo}}{6H_o^2\left(1+f_R(R_o)\right)}e^{-3(1+w_m)H_ot}\Big] \label{R27a}\\    \cr
 \delta(t)&=&k_1\Big[\frac{\rho_{mo}}{6H_0^2\left(1+f_R^b(R_o)\right)}\Big]\exp\Big[-3(1+w_m)H_ot+\frac{\rho_{mo}}{6H_o^2(1+f_R(R_o))}e^{-3(1+w_m)H_ot}\Big]   \label{R27b}                                                            \label{R28}
\end{eqnarray}
Also, the resolution of (\ref{R27}) and (\ref{R21}) give respectively:
\begin{eqnarray}
\delta_m(t)&=&k_1\exp\Big[\frac{\rho_{mo}}{6H_o^2}e^{-3(1+w_m)H_ot}\Big]  \label{R28a} \\   \cr
 \delta(t)&=&k_1\left(\frac{\rho_{mo}}{6H_0^2}\right)
\exp\Big[-3(1+w_m)H_ot+\frac{\rho_{mo}}{6H_o^2}e^{-3(1+w_m)H_ot}\Big]     \label{R28b}                                     
\end{eqnarray}
where $k_1$ is a integration constant.\\
The perturbation functions of the matter $\delta_m(t)$ (\ref{R27a}) and of the geometry $\delta(t)$ (\ref{R27b}) in above are defined for
$f_R^b(R_b)\neq-1$. We have the parameters  $w_m $ and $H_o$ that fulfill the following conditions: $w_m+1>0$ and $H_o>0$.\\ 
The solution $\delta_m(t)$ (\ref{R27a}) tends to $k_1$ when the time evolves . One take $k_1$ such as $0<k_1<1$. 
Then the model studied is stable for the ordinary matter with de Sitter solutions.\\With the function $\delta(t)$ (\ref{R27b}), one see that when $t\rightarrow \infty$, then $\delta \rightarrow 0$. Thus, in geometrical point of view, the model $f_\ast(R)$ is stable.\\ 
The perturbation functions $\delta_m(t)$ (\ref{R28a}) and $ \delta(t)$ (\ref{R28b}) tend respectively to the integration constant $k_1$ and to zero when the time evolves. We note an stability of the model $f_\ast(R)$ for the matter ($0<k_1<1$) and the geometry.
One find these different behaviors of the perturbation functions through the curved in the right-hand side of the fig2.\\ 
We see on the left-hand side (large $R$) that the both perturbations parameters decrease and tends to the finite values with the evolving time. The matter perturbation tends to $k_1$ (here, $k_1=0.5$) and the geometry's one is null with the time. Thus, the evolution of $\delta_m(t)$ depends on the integration constant $k_1$.\\
In the right-hand side (small $R$ ), the both perturbation functions decrease to the finite values, $k_1=0.5$ for the matter 
and zero for the geometry. Here, it's important to note that the function of the matter perturbation evolves depending on $k_1$ but the geometry's one doesn't.\\
Finally, one can say that the stability of $f_\ast(R)$ model for the de Sitter solutions depends on the values of the constant $k_1$ with the matter part, but it's stable on the geometrical side.

\subsection{Stability of power-law Solutions}
The scale factor for the power law solutions is written as:
\begin{eqnarray}
 a(t)\propto t^n\Longrightarrow H_b(t)=\frac{n}{t}                                                             \label{R32}
\end{eqnarray}
With the Hubble parameter expression, the relation (\ref{R17}) is written as:
\begin{eqnarray}
 \rho_{mb}(t)=\rho_{mo} e^{-3n(1+w_m)\ln t}                                                                   \label{R33}
\end{eqnarray}
The equations (\ref{R21}), (\ref{R22}) and (\ref{R33}) numerically give for $\beta=1$, $n=1$ and $\omega_m=0$, the following solutions:
\begin{eqnarray}
\delta_m(t)=k_2\exp\left[\frac{1}{2}\left(\frac{1}{t}+\frac{R_\ast\left(\sqrt{R_\ast-1}\arctan(\frac{\sqrt{R_\ast^2-1}}{\sqrt{6(R_\ast-1)}}t)-
\sqrt{R_\ast+1}\arctan(\frac{\sqrt{R_\ast^2-1}}{\sqrt{6(R_\ast+1)}}t)\right)}{2\sqrt{6(R_\ast^2-1)}}\right)\right]    \label{R33a}    \\ \cr
\delta(t)=k_2\frac{\rho_{mo}}{6\left(1-\left(\frac{R_\ast}{R_\ast+\frac{6}{t^2}}\right)^2\right)t}\exp\left[\frac{1}{2}\left(\frac{1}{t}+
\frac{R_\ast\left(\sqrt{R_\ast-1}\arctan(\frac{\sqrt{R_\ast^2-1}}{\sqrt{6(R_\ast-1)}}t)-\sqrt{R_\ast+1}\arctan(\frac{\sqrt{R_\ast^2-1}}{\sqrt{6(R_\ast+1)}}t)\right)}{2\sqrt{6(R_\ast^2-1)}}\right)\right]  \label{R33b}    
\end{eqnarray}
Now, one consider the equations  (\ref{R21}), (\ref{R23}) that give with (\ref{R33}), the following solutions :
\begin{eqnarray}
\delta_m(t)=k_3\exp\left[-\frac{t^{2-3n(1+\omega_m)}(1+\omega_m)}{2n(2-3n(1+\omega_m))}\right]   \label{R33c}  \\  \cr
\delta(t)=k_3\frac{\rho_{mo}t^{2-3n(1+\omega_m)}}{6n^2}\exp\left[-\frac{t^{2-3n(1+\omega_m)}(1+\omega_m)}{2n(2-3n(1+\omega_m))}\right] \label{R33d} 
\end{eqnarray} 
The perturbation functions (\ref{R33a}) and (\ref{R33b}) are defined for $R_\ast\neq1$. These functions decrease and tend respectively to a constant $k_2$ and to zero. Thus, by taking $0<k_2<1$, the both functions tend to zero and consequently, the model is stable. The representative graphs of these functions (fig 3 , left-hand side graphic(large $R$)) show better their behaviors. Here, we note that the matter density increases at the starting but decreases when the times evolves. Also, one see that the geometrical density is null. Thus, the $f_\ast(R)$ is stable for power law solutions in the case of large curvature.\\
The functions (\ref{R33c}) and (\ref{R33d}) is defined for $n\neq0$ and $2-3n(1+\omega_m)\neq0$, the perturbation density of the matter $\delta_m(t)$ increases at the starting and decreases when the time evolves and tends to $k_3$. Also  $\delta(t)$ tends to zero with the times . Observing these graphs (figure 3, right-hand side(small $R$ )), one see that the geometrical perturbation is firstly null. As we take $k_3$ such $0<k_3<1$, then the studied model is stable for the power law solutions in the small $R$ case. 
\newpage
 \begin{figure}[h]
 \centering
 \begin{tabular}{rl}
 \includegraphics[height=5.5cm,width=8cm]{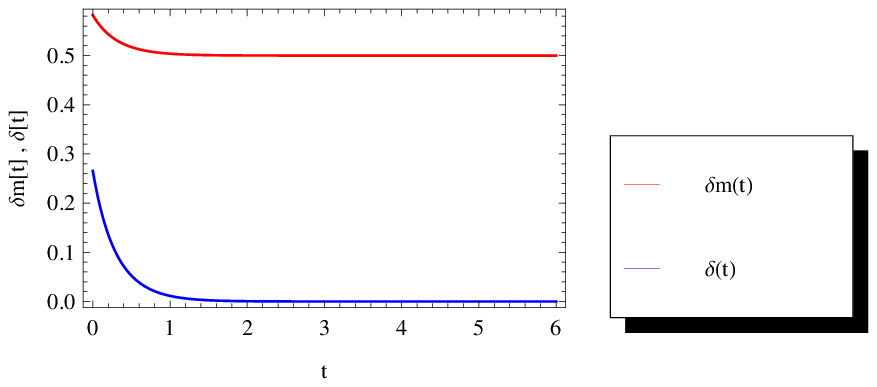}&
 \includegraphics[height=5.5cm,width=8cm]{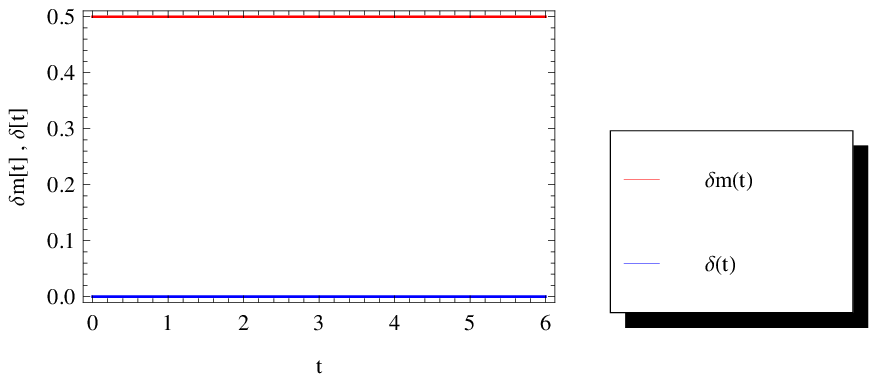} 
\end{tabular}
\caption{the graphs illustrate the convergence of the perturbation functions $\delta_m$, $\delta$ with the de Sitter solutions in the case of large $R$(left-hand side) and of small $R$(right-hand side). The graphs are plotted for $\rho_{mo}=1.8\times10^{-84}$, $R_\ast=1.004\times10^{-83}$ , 
$H_o=1.47\times10^{-42}$(small $R$ ) and $H_0=1$(large $R$ ).}
\label{fig1.}
\label{fig1}
\end{figure}

\begin{figure}[h]
 \centering
 \begin{tabular}{rl}
 \includegraphics[height=5.5cm,width=8cm]{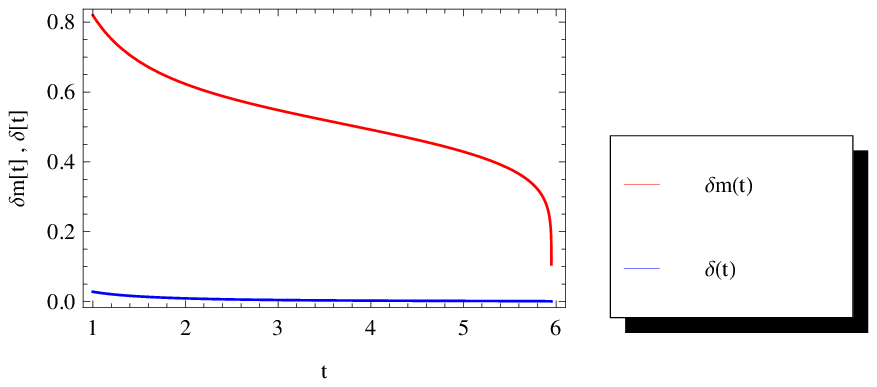}&
 \includegraphics[height=5.5cm,width=8cm]{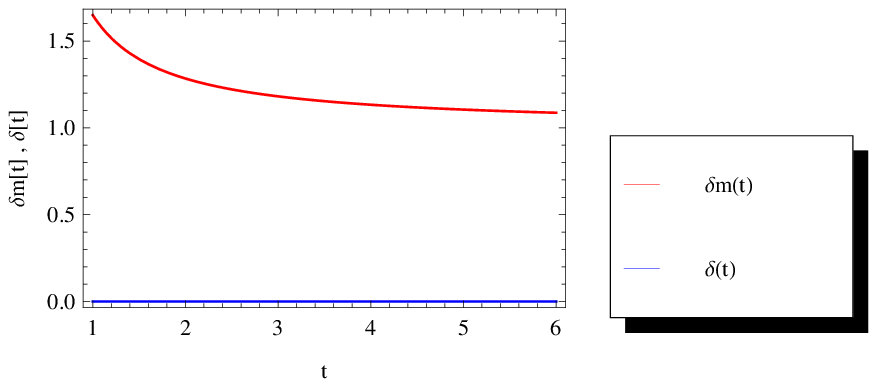} 
\end{tabular}
\caption{ the graphs illustrate the convergence of the perturbation functions $\delta_m$, $\delta$ with the power-law solutions in the case of large $R$(left-hand side) and of small $R$(right-hand side). The graphs are plotted for $\rho_{mo}=1.8\times10^{-84}$, $R_\ast=1.004\times10^{-83}$ , 
$H_o=1.47\times10^{-42}$(small $R$ ) and $H_o=1$(large $R$ ).}
\label{fig1}
\end{figure}

\newpage

\section{Conclusion}
In this work, we investigate the $R+f(R)$ modified gravity theory by exploring its dynamical analysis with application to gamma gravity model. 
This analysis shows six critical points that only one present the actual state of the universe(presence of both matter and the dominated 
dark energy). Each point describes a specific phase of space. There is no attractor fixed point but there are four conditional stable points
$(A_1,A_3,A_4,A_5)$. What it's important is that, we don't find a global repellor fixed point and the stability conditions of the conditional 
stable point  dependents on the parameters $n$, $u_3$, $u_2$ and $u_5$.\\

In the following, we study the stability of a dark energy model which behavior is the same with $f(R)$ $\gamma$-gravity model.
the generalization of models for different values of the parameter $\beta$  
The model is plotted and one can see the same behavior like the one of $f(R)$ $\gamma$-gravity  \cite{Rach18}. Note that with
this model General Relativity is recovered for large curvature. After that, we proceed to its stability analysis within de Sitter
and power law solutions for large and small curvature. We find that the model is more stable in the geometrical part than the
matter's one.

\vspace{0.5cm}
{\bf Acknowledgement:} R. D. Boko thanks MESRS of Benin government for partial financial support.

\end{document}